\documentclass[%
preprint,
amsmath,amssymb,
aps,
floatfix,
]{revtex4-1}

\usepackage{graphicx}
\usepackage{dcolumn}
\usepackage{bm}
\usepackage{color}
\usepackage{soul}

\usepackage{float} 

\begin{document}

\title{Attosecond control of spin polarization in electron-ion recollision driven by intense tailored fields}

\author{David Ayuso$^1$, Alvaro Jim\'enez-Gal\'an$^1$, Felipe Morales$^1$, Misha Ivanov$^{1,2,3}$ and Olga Smirnova$^{1,4}$}

\email{olga.smirnova@mbi-berlin.de}

\affiliation{$^1$Max-Born Institute for Nonlinear Optics and Short Pulse Spectroscopy, Max-Born-Stra{\ss}e 2A, D-12489 Berlin, Germany}
\affiliation{$^2$Department of Physics, Imperial College London, South Kensington Campus, SW72AZ London, UK}
\affiliation{$^3$Institute f\"ur Physik, Humboldt-Universität zu Berlin, Newtonstra{\ss}e 15, D-12489 Berlin, Germany}
\affiliation{$^4$Technische Universit\"at Berlin, Ernst-Ruska-Geb\"aude, Hardenbergstra{\ss}e 36A, 10623 Berlin, Germany}

\raggedbottom

\begin{abstract}
We show that electrons recolliding with the ionic core upon tunnel ionization of noble gas atoms driven by a strong circularly polarized laser field in combination with a counter-rotating second harmonic are spin polarized and that their degree of polarization depends strongly on the recollision time.
Spin polarization arises as a consequence of (1) entanglement between the recolliding electron and the ion, and (2) sensitivity of ionization to the sense of electron rotation in the initial state.
We demonstrate that one can engineer the degree of spin polarization as a function of time by tuning the relative intensities of the counter-rotating fields, opening the door for attosecond control of spin-resolved dynamics.
\end{abstract}

\maketitle

\section{Introduction}

The Stern-Gerlach experiment \cite{GerlachZP1922_1,GerlachZP1922_2,GerlachZP1922_3} revealed, in 1922, that an electron possesses an intrinsic angular momentum that is quantized and that is independent of its orbital angular momentum: the spin.
Electron spin governs the behavior of matter, arranging the electronic shells of the elements in the periodic table through the Pauli exclusion principle \cite{PauliZP1925} and giving rise to magnetism \cite{book_Coey_MagnetismAndMagneticMaterials}.
Ever since its discovery, finding ways of producing spin polarized electrons has attracted the interest of physicists \cite{book_Kessler_PolarizedElectrons}.
In 1969, Fano demonstrated that one-photon ionization of atoms with circularly polarized light in the energy region of Cooper minima can lead to the generation of electrons with a high degree of spin polarization \cite{FanoPRA1969}.
Another way of producing polarized currents is via ionization from a selected state of an atom or a molecule presenting fine structure splitting \cite{CherepkovJPB1981}.
This investigation has been extended to the multiphoton case in the perturbative regime \cite{LambropoulosAAMP1976,DixitPRA1981,NakajimaEPL2002}.
However, despite its importance, spin polarization with strong laser fields has received no attention until very recently \cite{BarthPRA2013_SpinPolarization,MilosevicPRA2016,HartungNatPhys2016,LiuPRA2016}.
The first theoretical predictions of spin polarization in noble gases upon strong field ionization with circularly polarized light \cite{BarthPRA2013_SpinPolarization} have just been experimentally confirmed \cite{HartungNatPhys2016}.

Spin polarization in the strong field regime is a consequence of electron-ion entanglement and the sensitivity of the ionization yield to the sense of electron rotation in the initial state \cite{BarthPRA2013_SpinPolarization}:
electrons that counter-rotate with the field ionize more easily than the co-rotating electrons, yielding different ionization rates for $p_-$ and $p_+$ electrons in noble gases \cite{BarthPRA2011,HerathPRL2012,BarthPRA2013_NonadiabaticTunneling,BarthPRA2013_NonadiabaticTunneling_comparison,BarthJPB2014_numerical,kaushal2015opportunities} and diatomic molecules \cite{LiuPRA2016}.
The possibility of inducing recollision of spin-polarized electrons with the parent ion can open new directions in attosecond spectroscopy \cite{HartungNatPhys2016,MilosevicPRA2016}.
Not surprisingly, the degree of spin polarization is higher for higher ellipticity of the ionizing field.
The flip side of the coin, however, is that high ellipticity of the ionizing field reduces the chance of electron return to the parent ion.
In this context, the use of an intense circularly polarized laser field in combination with its counter-rotating second harmonic, known as a bi-circular field, constitutes a powerful tool for introducing the spin degree of freedom into attosecond science, due to the opportunity to combine circular polarization with the efficient recollision offered by these fields \cite{EichmannPRA1995,LongPRA1995,MilosevicPRA2000,MilosevicOptLett2015,MedisauskasPRL2015,MilosevicPRA2016}.
The application of bi-circular fields can lead to the production of ultrashort circularly and elliptically polarized laser pulses in the XUV domain \cite{ZuoJNOPM1995,IvanovNatPhot2014,KfirNatPhot2015,MilosevicOptLett2015,MedisauskasPRL2015,MaugerJPB2016,BandraukJPB2016}.
Their chiral nature offers unique possibilities for probing molecular chirality \cite{SmirnovaJPB2015} or symmetry breaking \cite{BaykushevaPRL2016} at their natural time scales via high harmonic generation spectroscopy.
Recent theoretical work \cite{MilosevicPRA2016} has indicated that electrons produced upon strong field ionization with bi-circular fields are spin polarized.

Here we present a detailed theoretical study of spin polarization in electron-core recollision driven by bi-circular fields, emphasizing the possibilities for, and the physical mechanisms of controlling the degree of spin-polarization by changing the parameters of the bi-circular field.
The paper is organized as follows.
Section \ref{section_method} describes the theoretical approach, which is based on the strong field approximation (SFA).
Section \ref{section_results} describes our results, focusing on the analytical analysis of how the properties of the quantum electron trajectories define the spin polarization.
This allows us to establish the origin of spin polarization in bi-circular fields (section \ref{section_SPorigin}) and show how to achieve its attosecond control by tailoring the laser fields (section \ref{section_SPcontrol}).
Section \ref{section_conclusions} concludes the paper.

\section{Method} \label{section_method}

Consider ionization, followed by electron-parent ion recollision, of xenon atoms driven by a strong right circularly polarized (RCP) field in combination with the counter-rotating second harmonic.
The resulting electric field can be written, in the  
dipole approximation, as:
\begin{align}\label{eq_electricField}
\bold{F}(t) = \Big[ F_{0,\omega} \cos{(\omega t)} + F_{0,2\omega} \cos{(2\omega t)} \Big] \bold{\hat{x}} + \Big[ F_{0,\omega} \sin{(\omega t)} - F_{0,2\omega} \sin{(2\omega t)} \Big] \bold{\hat{y}}
\end{align}
where $F_{0,\omega}$ and $F_{0,2\omega}$ are the amplitudes of the right and left circularly polarized fields, respectively, with frequencies $\omega$ and $2\omega$.
Within the strong-field approximation (SFA), the continuum electron wave function at time $t$ is given by \cite{bookChapter_SmirnovaIvanov_AttosecondAndXUVPhysics}:
\begin{equation}\label{eq_SFAwf}
|\Psi(t)\rangle = i \int_{t_0}^{t}dt' e^{i\text{IP}(t'-t_0)} \bold{F}(t') \int d\bold{p} \text{ } \bold{d}(\bold{p}+\bold{A}(t')) \text{ } |\bold{p}+\bold{A}(t)\rangle_V 
\end{equation}
where IP is the ionization potential,
$\bold{p}$ is the drift (canonical) momentum, related to the the kinetic momentum $\bold{k}(t)$ by $\bold{k}(t)=\bold{p}+\bold{A}(t)$, 
$\bold{d}(\bold{p}+\bold{A}(t)) = \langle\bold{p}+\bold{A}(t)|\hat{\bold{d}}|\Psi_0\rangle$ is the transition dipole matrix element from the initial ground state $|\Psi_0\rangle$ (the system is assumed to be in the ground state at $t=t_0$) 
to a Volkov state $|\bold{p}+\bold{A}(t)\rangle_V$, given by
\begin{equation}\label{eq_VolkovState}
|\bold{p}+\bold{A}(t)\rangle_V = \frac{1}{(2\pi)^{3/2}} e^{-iS_V(t,t',\bold{p})} e^{i[\bold{p}+\bold{A}(t)]\cdot\bold{r}}
\end{equation}
where $S_V(t,t',\bold{p})$ is the Volkov phase:
\begin{equation}\label{eq_VolkovPhase}
S_V(t,t',\bold{p}) = \frac{1}{2} \int_{t'}^{t} d\tau \big[\bold{p}+\bold{A}(\tau)\big]^2
\end{equation}
Eq. \ref{eq_SFAwf} can be used to calculate different observables, such as photoelectron yields, induced polarization and harmonic spectra \cite{bookChapter_SmirnovaIvanov_AttosecondAndXUVPhysics}.
Here we are interested in analyzing the degree of spin-polarization of the electrons that are driven back to the ionic core.
This requires a measure of the recollision probability, resolved on the state of the ion and on the spin of the returning electron.
The latter is determined by the initial magnetic quantum number of the state from which the electron tunnels and the state of
the ion that has been created upon ionization, as described in \cite{BarthPRA2013_SpinPolarization}.
As for the recollision probability, given that the size of the returning wave packet far exceeds the size of the atom, an excellent measure of the recollision amplitude is the projection of the continuum wave function (eq. \ref{eq_SFAwf}) $|\Psi(t)\rangle$ on any compact object at the 
origin; the recollision current will
scale with the object area. To obtain the required recollision
probability density at the origin, we simply project $|\Psi(t)\rangle$ on the delta-function  at the origin, yielding
\begin{equation}\label{eq_SFAwf1}
a_{\rm rec}(t)= i \int_{t_0}^{t}dt' \bold{F}(t') \int d\bold{p} \text{ } \bold{d}(\bold{p}+\bold{A}(t')) \text{ } e^{-i[S_V(t,t',\bold{p})+\text{IP}(t-t')]} \text{ } 
\end{equation}

The degree of spin polarization of the recolliding electrons as a function of the recollision time $t$ is given by the normalized difference between the recollision probability densities for electrons recolliding with spin up ($w_{\uparrow}(t)=|a_{\uparrow}(t)|^2$) and spin down ($w_{\downarrow}(t)=|a_{\downarrow}(t)|^2$) \cite{BarthPRA2013_SpinPolarization}:
\begin{equation}\label{eq_spinPol}
\text{SP}(t) = \frac{w_{\uparrow}(t)-w_{\downarrow}(t)}{w_{\uparrow}(t)
+w_{\downarrow}(t)} 
\end{equation}
The densities $w_{\uparrow}(t)$ and $w_{\downarrow}(t)$ are obtained from the recollision densities $w^{p_+,p_-}_{\text{IP}^{^2P_{3/2,1/2}}}(t)=\big|a^{p_+,p_-}_{\text{IP}^{^2P_{3/2,1/2}}}(t)\big|^2$ correlated to ionization from the $p_+$ and $p_-$ orbitals,
resolved on the ionic states $^2P_{3/2}$ and $^2P_{1/2}$, and the corresponding Clebsch-Gordan coefficients 
\cite{BarthPRA2013_SpinPolarization}:
\begin{align}
w_{\uparrow}(t)   &= w^{p_+}_{\text{IP}^{^2P_{3/2}}}(t) + \frac{2}{3} w^{p_-}_{\text{IP}^{^2P_{1/2}}}(t) + \frac{1}{3} w^{p_-}_{\text{IP}^{^2P_{3/2}}}(t) \label{eq_ionizationRate_spinUP} \\
w_{\downarrow}(t) &= w^{p_-}_{\text{IP}^{^2P_{3/2}}}(t) + \frac{2}{3} w^{p_+}_{\text{IP}^{^2P_{1/2}}}(t) + \frac{1}{3} w^{p_+}_{\text{IP}^{^2P_{3/2}}}(t) \label{eq_ionizationRate_spinDOWN}
\end{align}
The contribution of the $p_0$ orbital is negligible \cite{BarthPRA2011,BarthPRA2013_NonadiabaticTunneling}.
The key quantities in these expressions are the recollision densities resolved on the initial orbital and the final ionic state, $w^{p_-}_{\text{IP}^{^2P_{1/2}}}=\big|a^{p_-}_{\text{IP}^{^2P_{1/2}}}\big|^2$, etc. 
Application of the saddle-point method (see e.g. \cite{bookChapter_SmirnovaIvanov_AttosecondAndXUVPhysics}) to the integral eq. \ref{eq_SFAwf1} allows us to perform the semi-classical analysis of this expression in terms of electron trajectories, getting insight into the physical origin of spin polarization during recollision.
The saddle points are calculated by solving the following set of equations \cite{bookChapter_SmirnovaIvanov_AttosecondAndXUVPhysics}:
\begin{align}
\frac{[\bold{p}+\bold{A}(t_i)]^2}{2}+\text{IP} &= 0    
\label{eq_SP_tunneling} \\
\int_{t_i}^{t_r} d\tau [\bold{p}+\bold{A}(\tau)] &= 0   \label{eq_SP_zeroDisplacement} 
\end{align}
where 
IP is the ionization potential, $t_i$ and $t_r$ are the 
complex ionization and recollision times, respectively.
Eq. \ref{eq_SP_tunneling} describes tunneling and eq. \ref{eq_SP_zeroDisplacement} requires that the electron returns to the core.

\begin{figure}[H]
\centering
\includegraphics[scale=0.5]{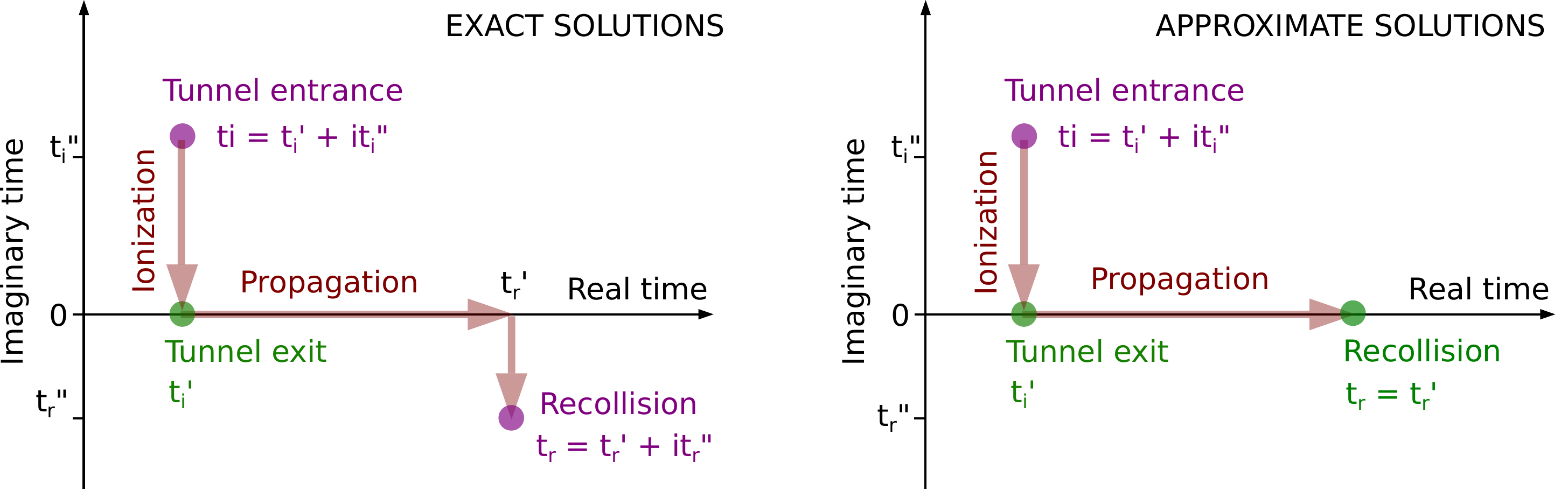}
\caption{Schematic representation of the contour time integration of the action.
Ionization starts at a complex time $t_i = t_i' + it_i''$, the electron tunnels out of the potential barrier at the real time $t_i'$, and returns to the ionic core at $t_r = t_r' + it_r''$ (left panel).
If the imaginary part of the recollision time is sufficiently small, one can keep the recollision time on the real time axis (right panel), simplifying the treatment of the cutoff region.}
\label{fig_contourTimeIntegration}
\end{figure}

Fig. \ref{fig_contourTimeIntegration} shows a schematic representation of the process on the complex time plane.
The electron enters the barrier at complex time $t_i = t_i' + it_i''$.
The motion in the classically forbidden region occurs along the imaginary time axis and the electron is born in the continuum at a real time $t_i'$.
As a result, the recollision time $t_r$ and the canonical momentum $\bold{p}$ are, in general, complex.
To further simplify the analysis, we can take into account that for most of the relevant trajectories the imaginary part of their recollision time is rather small.
This allows one to keep the recollision time on the real time axis, also simplifying the treatment of the usual divergences near 
the cutoff region, see \cite{bookChapter_SmirnovaIvanov_AttosecondAndXUVPhysics}.

The recollision densities correlated to ionization from $p_+$ and $p_-$ orbitals are proportional to:
\begin{equation}\label{eq_ionizationRate}
w^{p_m}_{\text{IP}} \propto \Big|e^{-i[S_V(t_r,t_i,\bold{p})+\text{IP}(t_r-t_i)] + im\phi_{\bold{k}(t_i)}} \Big|^2
\simeq e^{2\Im{\{S_V(t_i',t_i,\bold{p})\}}-2\text{IP}t_i''} \text{ } e^{-2m\Im{\{\phi_{\bold{k}(t_i)}}\}}
\end{equation}

In this expression, the first key quantity that determines the magnitude of $w_{\text{IP}}^{p_m}$ is the imaginary part of action. 
It is mostly accumulated between the times $t_i=t_i'+it_i''$ and $t_i'$, i.e. in the classically forbidden region.
The second key quantity, which depends on the projection $m$ of the angular momentum, is the complex-valued ionization angle $\phi_{\bold{k}(t_i)}$.
It is given by the following expression:
\begin{equation}\label{eq_ionizationAngle}
\phi_{\bold{k}(t_i)} = \text{atan}\Bigg(-\frac{k_x'(t_i)}{k_y'(t_i)}\Bigg) + i \text{ atanh}\Bigg(\frac{k_x'(t_i)}{k_y''(t_i)}\Bigg) 
\end{equation}
with $k_x(t_i)=k_x'(t_i)+ik_x''(t_i)$ and $k_y(t_i)=k_y'(t_i)+ik_y''(t_i)$ being the complex velocities along x and y directions, respectively.
Note that the difference between the recollision densities from $p_+$ and $p_-$ orbitals depends solely on the imaginary part of the ionization angle.

Finally, the electron recollision energy is calculated as
\begin{equation}
E_{\text{rec}} = \frac{[\bold{p}+\boldsymbol{A}(t_r)]^2}{2} \label{eq_SP_recollisionEnergy}
\end{equation}
neglecting small imaginary contribution when keeping $t_r$ on the real time axis.

\section{Results} \label{section_results}

\begin{figure}[H]
\centering
\includegraphics[scale=0.2]{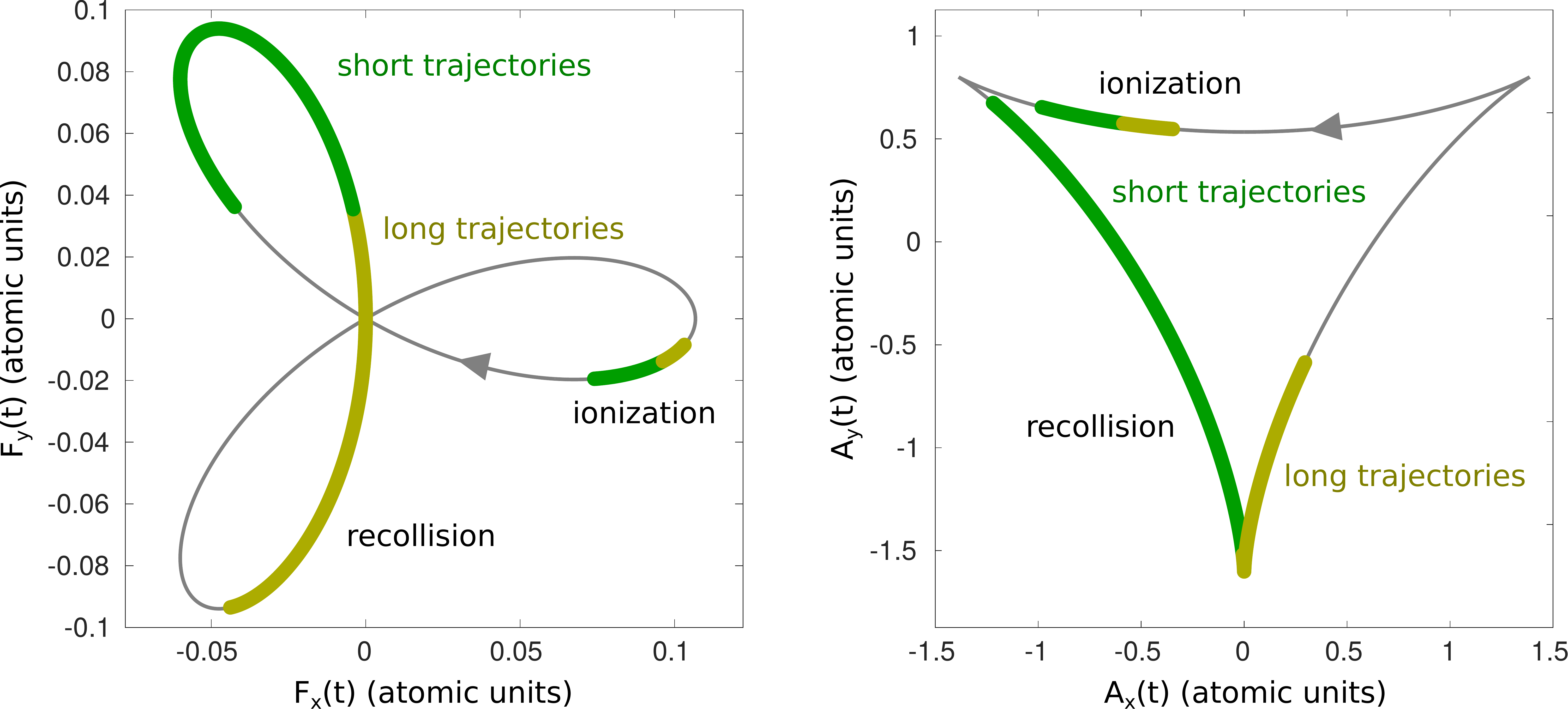}
\caption{Electric field (left panel) and vector potential (right panel) resulting from combining a RCP field of frequency $\omega = 0.05$ a.u. and intensity $I=10^{14}$ W cm$^{-2}$ with a LCP field of frequency $2\omega$ and equal intensity.
The ionization and recollision time-windows are indicated in the figures for short (green) and long (yellow) trajectories for one of the three ionization bursts.}
\label{fig_field}
\end{figure}

The Lissajous curves of the electric field considered here (see eq. \ref{eq_electricField}) and of the corresponding vector potential $\bold{A}(t)$, given by $\bold{F}(t)=-d\bold{A}(t)/dt$, are shown in fig. \ref{fig_field}, as well as the ionization and the recollision time windows (the field parameters are given in the fig. \ref{fig_field} caption).
The resulting electric field has a three-fold symmetry, with 3 peaks per cycle oriented at angles $0$, $2\pi/3$ and $4\pi/3$ rad in the $xy$ plane.
Ionization is more likely to occur near the maxima of the electric field, where the tunneling barrier is thinner.
Electrons liberated just before these maxima are unlikely to return to the core, those released after the maximum can recollide.

\begin{figure}[H]
\centering
\includegraphics[scale=0.2]{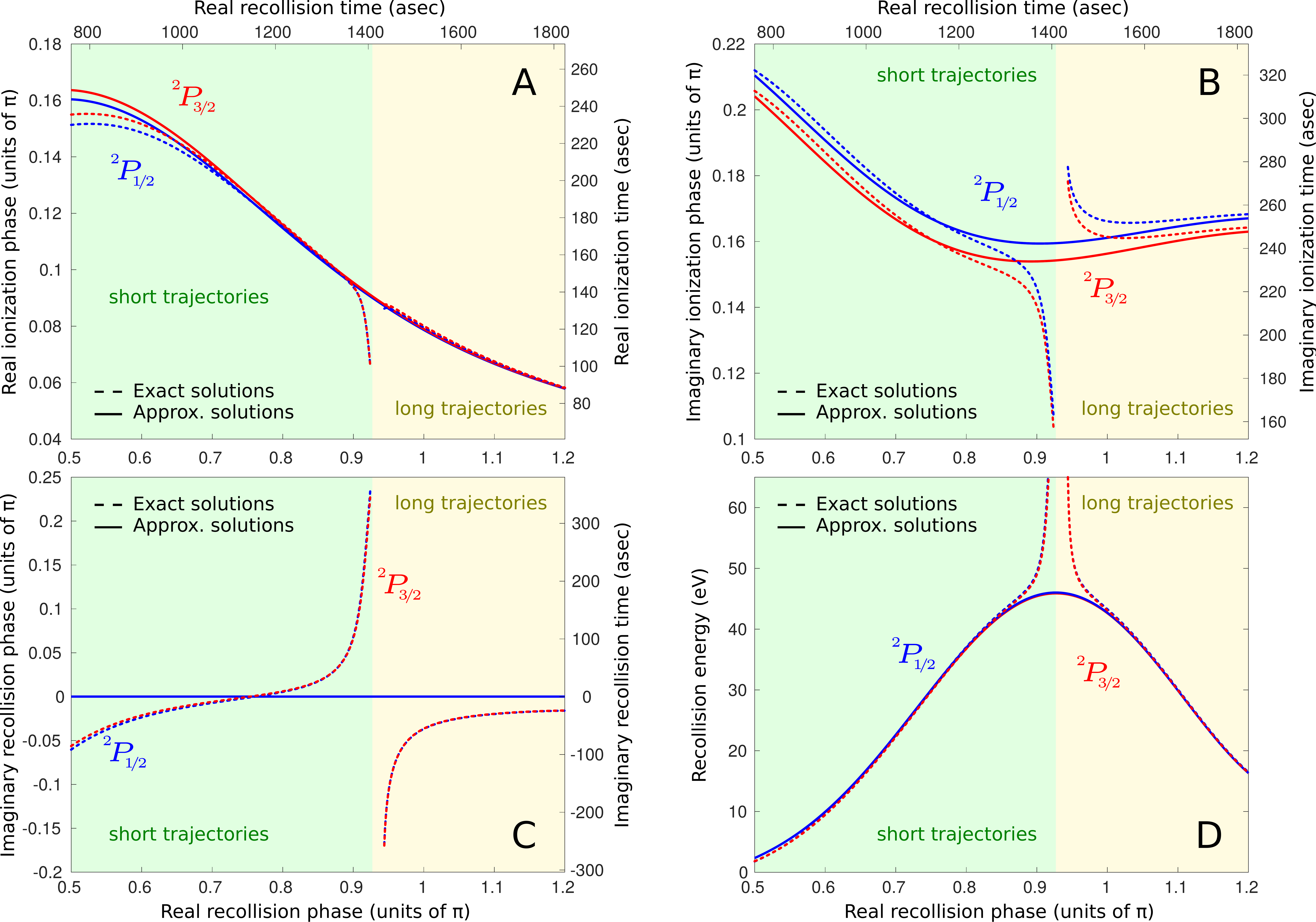}
\caption{Saddle point solutions for the bi-circular field represented in fig. \ref{fig_field} as functions of the real part of the recollision time: real (A) and imaginary (B) parts of the ionization time, imaginary part of the recollision time (C), and recollision energy (D).
Full saddle points (dashed lines) have been calculated allowing both ionization and recollision times to be complex, whereas approximate solutions (full lines) have been obtained by keeping the time of return on the real time axis (see fig. \ref{fig_contourTimeIntegration}).
Results are shown for the ionic states of xenon $^2P_{3/2}$ (red lines) and $^2P_{1/2}$ (blue lines), with ionization potentials IP$^{^2P_{3/2}} = 12.13$ eV and IP$^{^2P_{1/2}} = 13.43$ eV.}
\label{fig_saddlePoints}
\end{figure}

Consider strong field ionization of a xenon atom from the outermost 5p shell.
The spin-orbit interaction splits the energy levels of the ion into $^2P_{3/2}$ and $^2P_{1/2}$, with ionization potentials IP$^{^2P_{3/2}} = 12.13$ eV and IP$^{^2P_{1/2}} = 13.43$ eV.
Our calculations considered both ionic states, as needed for calculating spin polarization.
The saddle point equations (eqs. \ref{eq_SP_tunneling} and \ref{eq_SP_zeroDisplacement}) have been solved numerically, allowing the ionization and return times to be complex (exact solutions), and also by keeping the return time on the real time axis (approximate solutions), as represented in fig. \ref{fig_contourTimeIntegration}.
The real and imaginary parts of the ionization time, the complex part of the recollision time and the recollision energy (evaluated using eq. \ref{eq_SP_recollisionEnergy}) are shown in fig. \ref{fig_saddlePoints}, as functions of the real part of the return time.
Our exact solutions agree with those reported previously in \cite{MilosevicPRA2000} and the approximate solutions agree well with the exact ones.
We can see that the imaginary part of the recollision time (fig. \ref{fig_saddlePoints}c) is rather small, except near the cutoff, where the saddle point method diverges.
The main advantage of using approximate solutions and keeping the recollision time on the real time axis is that the ionization time and the recollision energy behave smoothly in the vicinity of the cutoff, while being very similar to the exact solutions outside this region.

Let us compare now the results for the states $^2P_{3/2}$ and $^2P_{1/2}$ of the ion.
As expected, the real part of the ionization time (fig. \ref{fig_saddlePoints}a) and the recollision energy (fig. \ref{fig_saddlePoints}c) are almost identical in both cases.
The imaginary part of the ionization time, however (fig. \ref{fig_saddlePoints}b), is slightly smaller for the $^2P_{3/2}$ state, with the lower IP, resulting in higher ionization amplitudes.\\


\begin{figure}[H]
\centering
\includegraphics[scale=0.2]{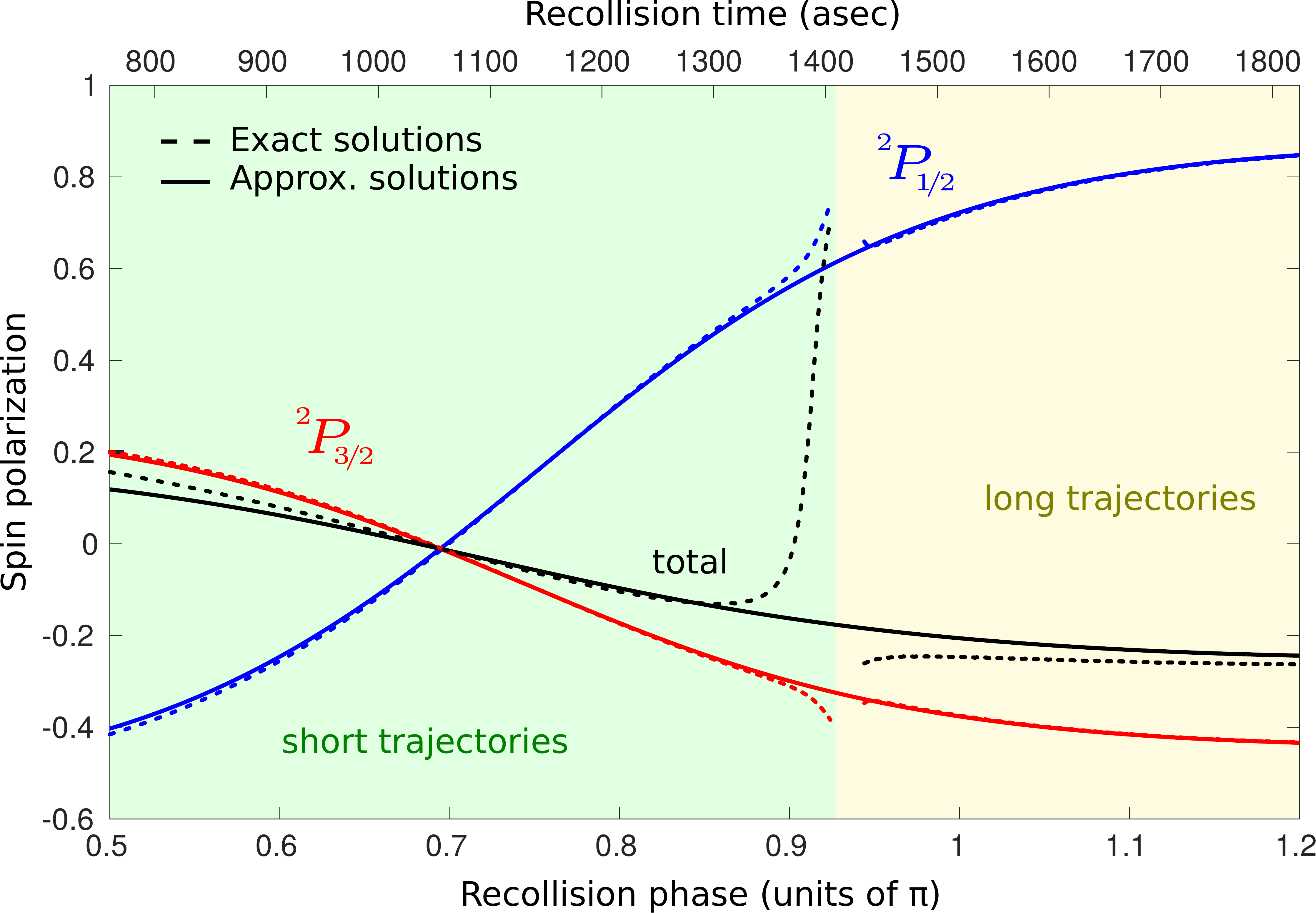}
\caption{Total spin polarization (black lines) and spin polarization resolved in the $^2P_{3/2}$ (red lines) and in the $^2P_{1/2}$ (blue lines) states of the core as a function of the recollision time.
Spin polarization has been calculated using the exact (full lines) and the approximate (dashed lines) saddle points solutions shown in fig. \ref{fig_saddlePoints}.}
\label{fig_SPvstr}
\end{figure}

We have evaluated the degree of spin polarization in recollision (eq. \ref{eq_spinPol}) using the saddle point solutions shown in fig. \ref{fig_saddlePoints}.
Total spin polarization is shown in fig. \ref{fig_SPvstr} as a function of the recollision time, together with the degree of polarization resolved in the $^2P_{1/2}$ and $^2P_{3/2}$ states of the core.
It is clear from the figure that recolliding electrons are spin-polarized and that their degree of polarization depends strongly on the recollision time.
Electrons that return to the core at earlier (later) times are more likely to have spin up (down).
Note also that spin polarization resolved in the ionic states $^2P_{1/2}$ and $^2P_{3/2}$ has opposite sign.
Both spin polarization resolved on the states of the ion and the 
total spin polarization change sign at the recollision phase (time) 
of $0.7\pi$ rad ($1.11$ fsec).
Each return time is associated with a given recollision energy, which is the well-known time-energy mapping \cite{bookChapter_SmirnovaIvanov_AttosecondAndXUVPhysics} (see fig. \ref{fig_saddlePoints}d).
Fig. \ref{fig_SPvstr} shows spin polarization as a function of the recollision energy for short and long trajectories.
Whereas for the short trajectories spin polarization changes dramatically as a function of the recollision energy, for the long trajectories the variation is rather smooth. 

\begin{figure}[H]
\centering
\includegraphics[scale=0.2]{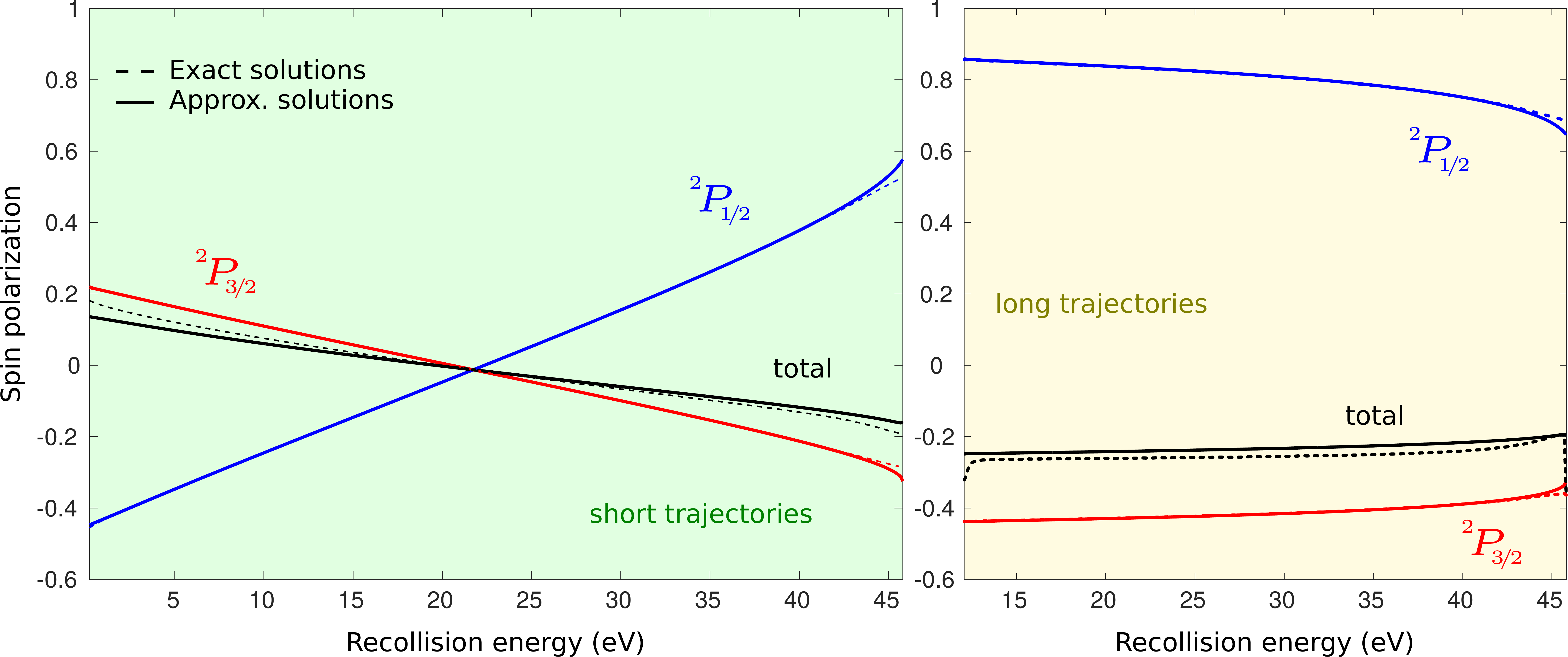}
\caption{Total spin polarization (black lines) and spin polarization resolved in the $^2P_{3/2}$ (red lines) and in the $^2P_{1/2}$ (blue lines) states of the core as a function of the recollision energy for short (left panel) and long (right panel) trajectories.
Spin polarization has been calculated using the exact (full lines) and the approximate (dashed lines) saddle points solutions shown in fig. \ref{fig_saddlePoints}.}
\label{fig_SPvsE}
\end{figure}

\subsection{Origin of spin polarization} \label{section_SPorigin}

\begin{figure}[H]
\centering
\includegraphics[scale=0.2]{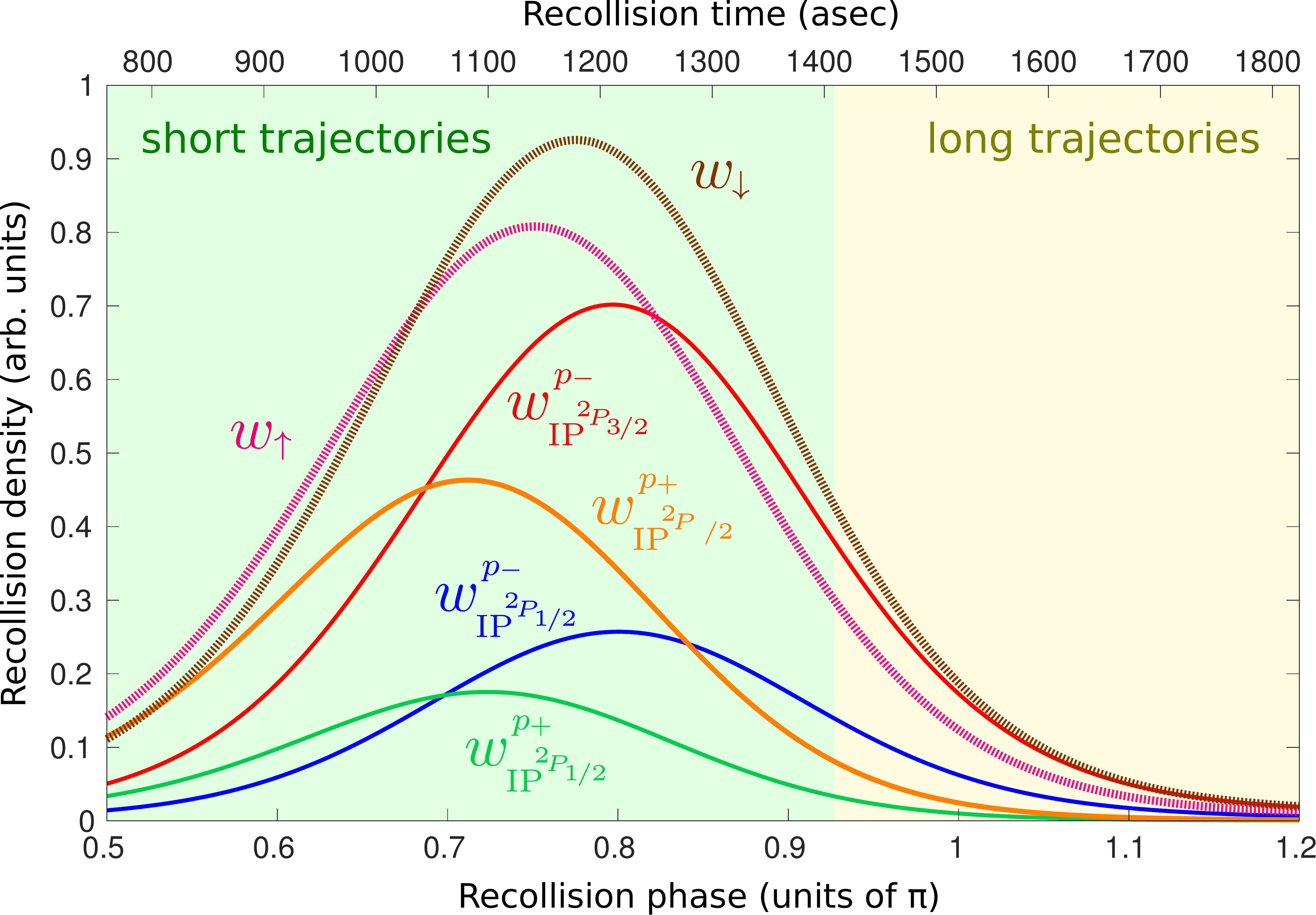}
\caption{Recollision densities for $p_+$ and $p_-$ electrons correlated to the states of the ion $^2P_{3/2}$ and $^2P_{1/2}$ as a function of the recollision time (full lines) and total recollision densities for electrons with spin up and spin down (dashed lines), calculated using the approximate quantum orbits resulting from keeping the time of return on the real time axis.}
\label{fig_recollisionDensity}
\end{figure}

To better understand the physical origin of spin polarization in recollision, let us analyze the recollision densities for different ionic channels. 
These are presented in fig. \ref{fig_recollisionDensity} as a function of the recollision time, as well as the total recollision densities corresponding to electrons with spin up and spin down (eqs. \ref{eq_ionizationRate_spinUP} and \ref{eq_ionizationRate_spinDOWN}).
There are three important things worth noting here.
First, the recollision densities correlated to the $^2P_{3/2}$ state of the core ($w^{p_-}_{\text{IP}^{^2P_{3/2}}}$ and $w^{p_+}_{\text{IP}^{^2P_{3/2}}}$) are higher than those for the $^2P_{1/2}$ state ($w^{p_-}_{\text{IP}^{^2P_{1/2}}}$ and $w^{p_+}_{\text{IP}^{^2P_{1/2}}}$) because the 
lower ionization potential of this ionic state leads to smaller imaginary ionization times (see fig. \ref{fig_saddlePoints}b) - the tunneling barrier is thinner.
Second, all recollision densities exhibit a maximum value that arises at lower recollision times in the case of the $p_+$ orbital ($w^{p_+}_{\text{IP}^{^2P_{3/2}}}$ and $w^{p_+}_{\text{IP}^{^2P_{1/2}}}$).
Third, the densities resolved on the $^2P_{3/2}$ and $^2P_{1/2}$ states of the core cross at $\phi_r=0.69\pi$ rad ($t_r=1044$ asec) and $\phi_r=0.70$ rad ($t_r=1061$ asec), respectively, leading to changes of sign in spin polarization (see fig. \ref{fig_SPvstr}).

\begin{figure}[H]
\centering
\includegraphics[scale=0.15]{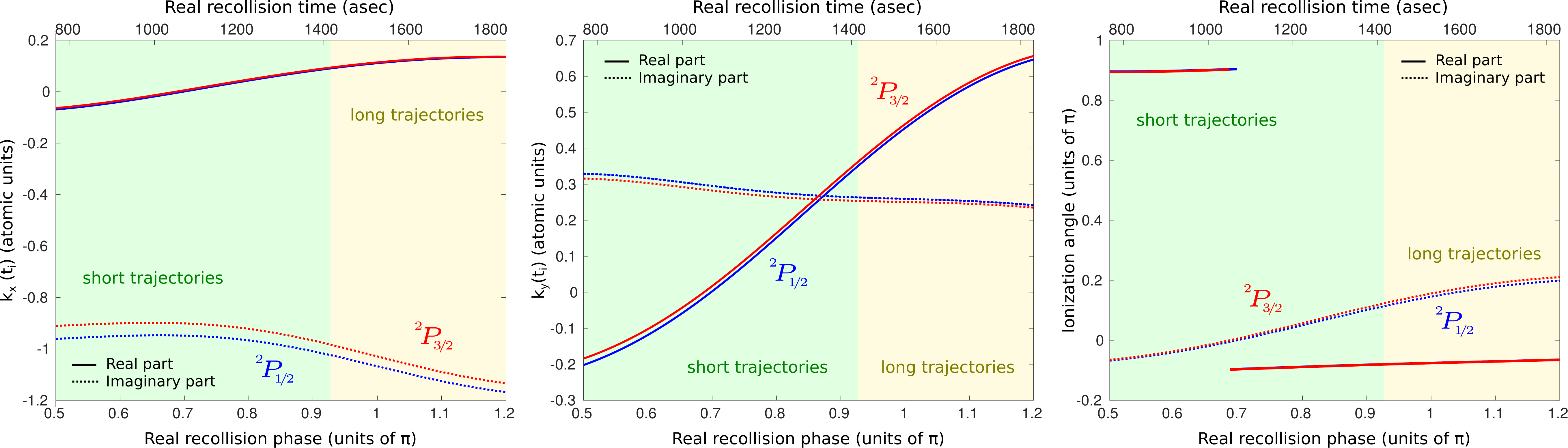}
\caption{Real (full lines) and imaginary (dashed lines) parts of the electron velocity at $t=t_i$ (when it enters the classically forbidden region) along the x (left panel) and the y (central panel) directions, and ionization angle (right panel) calculated using eq. \ref{eq_ionizationAngle}, as a function of the recollision time.}
\label{fig_ionizationVelocity}
\end{figure}

In order to understand these features, we have examined the saddle point solutions at $t=t_i$, when the electron enters the classically forbidden region.
The ionization velocity and the ionization angle are shown in fig. \ref{fig_ionizationVelocity} as a function of the recollision time.
We can see that, for a recollision phase (time) of $0.7\pi$ rad ($1.11$ fsec), the real part of the ionization angle presents a jump of $\pi$ and its imaginary component becomes zero.
A purely real ionization angle leads to equal tunnelling probabilities for $p_+$ and $p_-$ orbitals (see eq. \ref{eq_ionizationRate}) and thus no spin polarization.

\begin{figure}[H]
\centering
\includegraphics[scale=0.2]{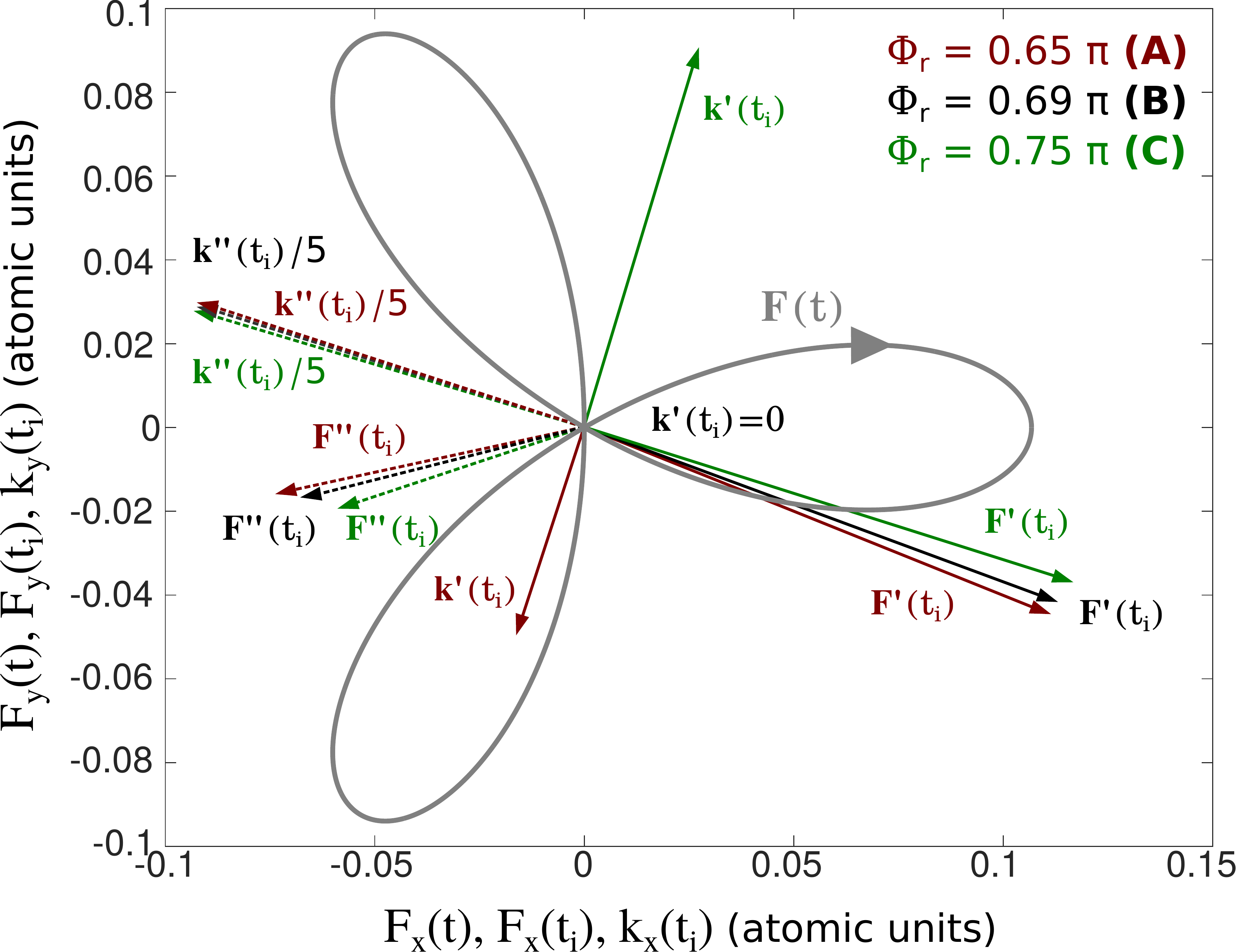}
\caption{Real and imaginary parts of the kinetic momentum $\bold{k}(t_i)$ and the electric field $\bold{F}(t_i)$ at the saddle point of ionization $t=t_i$;
$\bold{k}(t_i)=\bold{k}'(t_i)+i\bold{k}''(t_i)$ and $\bold{F}(t_i)=\bold{F}'(t_i)+i\bold{F}''(t_i)$.
Solutions are shown for one ionization burst.
The electric field considered here, resulting from combining a RCP field of frequency $\omega = 0.05$ a.u. and intensity $I=10^{14}$ W cm$^{-2}$ with a counter-rotating second harmonic of equal intensity, is represented in the figure.}
\label{fig_field+kti+Fti}
\end{figure}

\begin{figure}[H]
\centering
\includegraphics[scale=0.2]{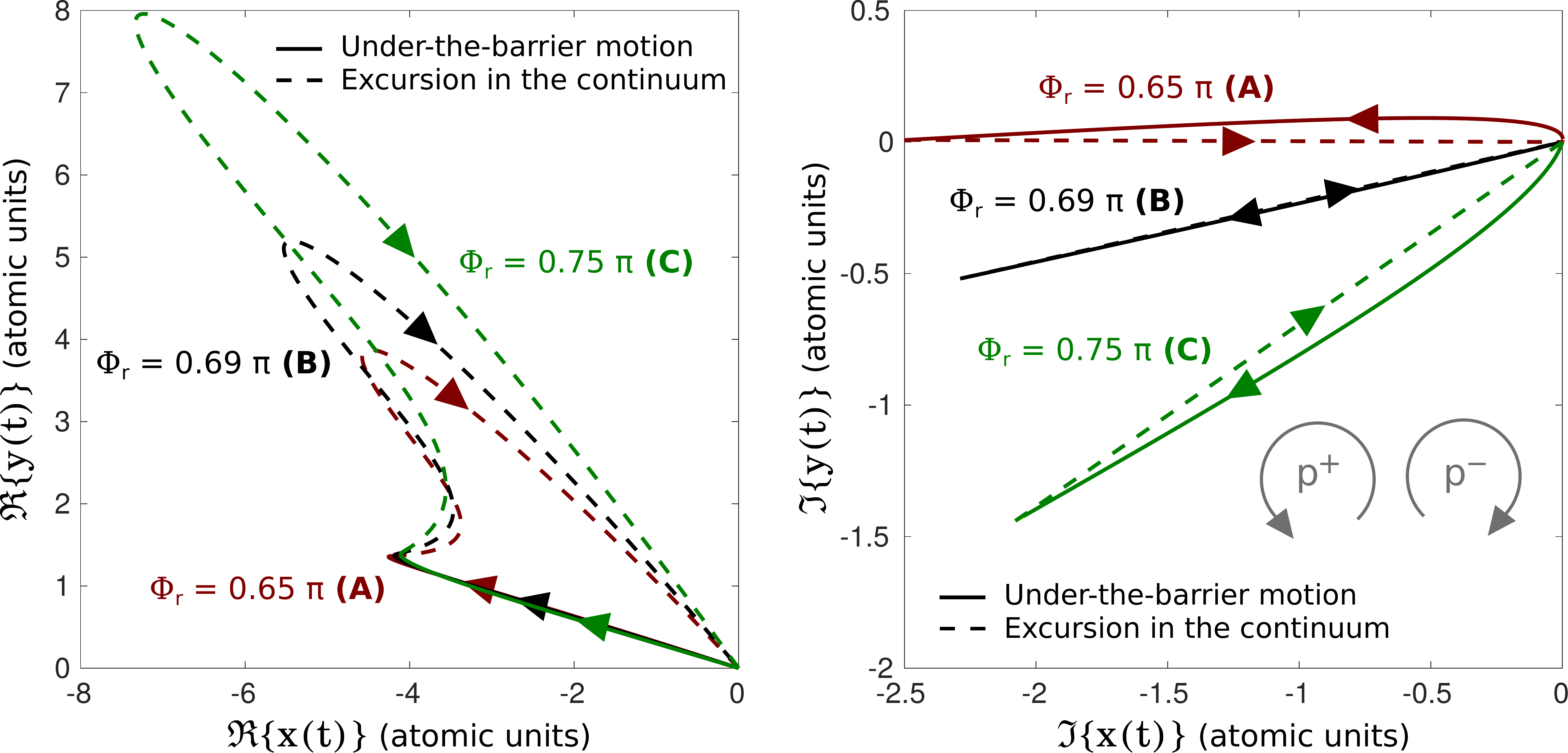}
\caption{Real (left panel) and imaginary (right panel) components of the quantum orbits that recollide with the core at $\phi_r=0.65\pi$ (traj. A, red lines), $0.69\pi$ (traj. B, black lines) and $0.75\pi$ (traj. C, green lines).
The corresponding recollision times ($t_r=\phi_r/\omega$) are $t_r=314$, $334$ and $363$ asec.
For illustration purposes, the sense of rotation of electrons in $p_+$ and $p_-$ orbitals is depicted in the right panel.}
\label{fig_trajectories}
\end{figure}

The time-dependent sensitivity of the recollision densities to the sense of rotation of the electron in its initial state can be understood by examining different quantum trajectories. 
Fig. \ref{fig_field+kti+Fti} contains a representation of the values of the electric field and the ionization velocity at $t=t_i$ of three quantum orbits that recollide with the $^2P_{3/2}$ state of the ion at different times: $\phi_r=0.65\pi$ rad (positive spin polarization), $\phi_r=0.69\pi$ rad (no spin polarization) and $\phi_r=0.75\pi$ rad (negative spin polarization), calculated by keeping the time of return on the real time axis.
We will refer to them as trajectories A, B, and C, respectively.
The three trajectories have similar values of $\bold{k}''(t_i)$ and $\bold{F}(t_i)$.
However, their values of $\bold{k}'(t_i)$ are very different. 
Let us analyze the motion of the electron through the classically forbidden region, which occurs in imaginary time (see fig. \ref{fig_contourTimeIntegration}) and along the complex plane of spatial coordinates ($\bold{r}=\bold{r}'+i\bold{r}''$).
The real part of the trajectory depends on $\bold{k}''$ and $\bold{F}'$ according to $\bold{k}''=d\bold{r}'/d\tau$ and $\bold{F}'=d\bold{k}''/d\tau$, with $\tau$ being the complex time variable.
Under the barrier, $d\tau=-dt''$ (see fig. \ref{fig_contourTimeIntegration}).
Equivalently, the motion in the plane of imaginary coordinates is dictated by $\bold{k}'=-d\bold{r}''/d\tau$ and $\bold{F}''=-d\bold{k}'/d\tau$.
Trajectories A, B and C are depicted in fig. \ref{fig_trajectories}.
Their real parts in the classically forbidden region are almost identical because they present similar values of $\bold{k}''(t_i)$ and $\bold{F}'(t_i)$. 
The motion in the imaginary plane, however, is different due to the very distinct values of $\bold{k}''(t_i)$.
Trajectory B presents $\bold{k}'(t_i)=0$ and thus its motion in the complex plane is solely dictated by the imaginary value of the electric field, which barely changes its direction during tunneling.
Thus, the motion in the imaginary plane occurs along a straight line.
The initial values of $\bold{k}'$ for trajectories A and C are non zero and point in opposite directions (see fig. \ref{fig_field+kti+Fti}).
During tunneling, they are modified by $\bold{F}''$, giving rise to clockwise motion in trajectory A and to anti-clockwise motion in trajectory B along the plane of imaginary coordinates (see fig. \ref{fig_trajectories}).
Because of its initial angular momentum, $p_+$ ($p_-$) electrons can be driven more easily along trajectory A (B) than $p_-$ ($p_+$) electrons, which leads to different recollision densities and leads to the time-dependent spin-polarization in recollision.

\subsection{Attosecond control of spin polarization} \label{section_SPcontrol}

\begin{figure}[H]
\centering
\includegraphics[scale=0.16]{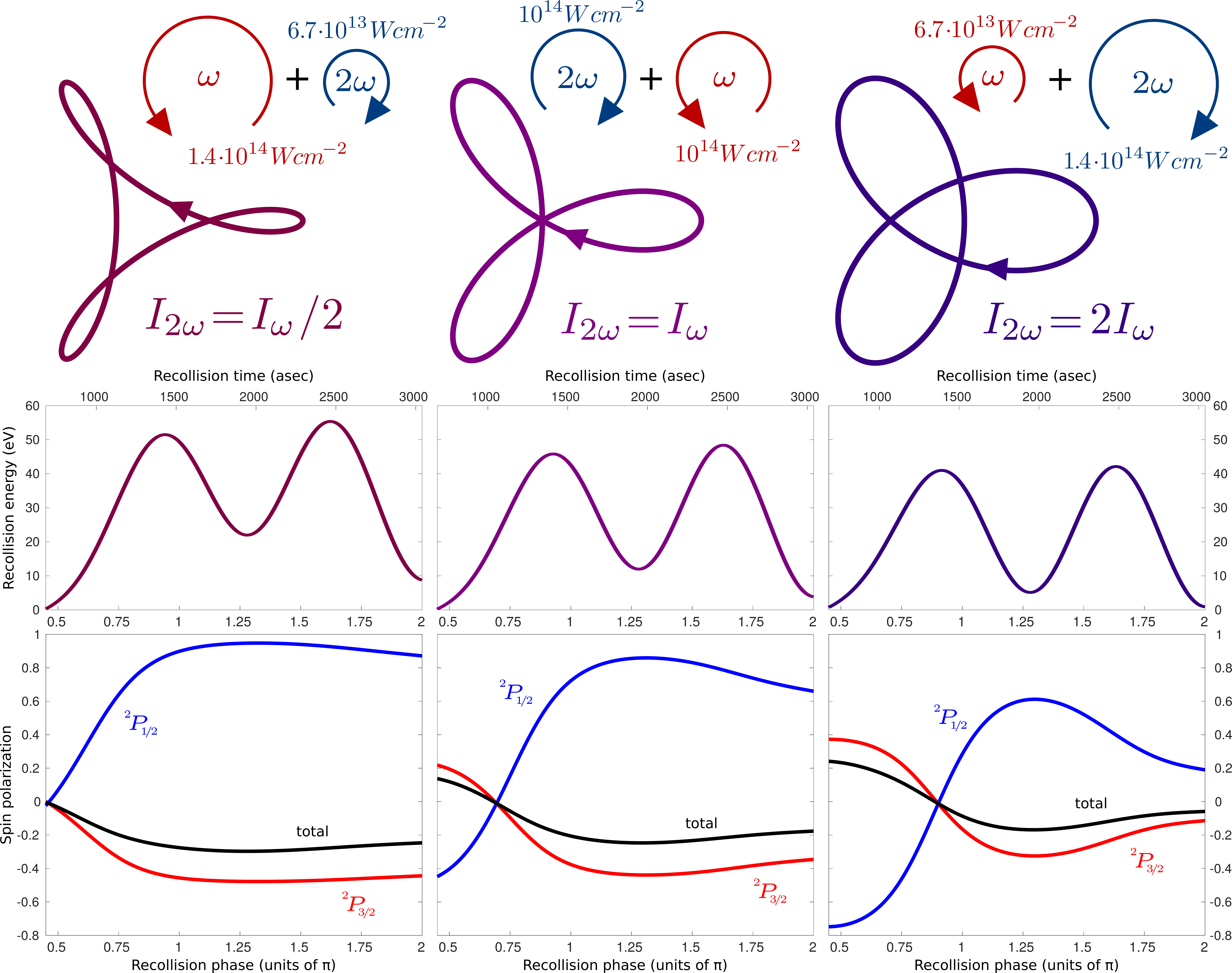}
\caption{Attosecond control of spin polarization.
Upper figures: Lissajous curves representing the electric fields resulting from combining a RCP field with frequency $\omega=0.05$ a.u. and a LCP field with frequency $2\omega$ with different relative intensities:
$I_{2\omega}=I_{\omega}/2$ (left column), $I_{2\omega}=I_{\omega}$ (central column) and  $I_{2\omega}=2I_{\omega}/2$ (right column).
The values of $I_{\omega}$ and $I_{2\omega}$ considered in each case are indicated in the figure.
Middle panels: recollision energy as a function of the recollision time.
Lower panels: spin polarization as a function of the recollision time.
Results have been calculated by keeping the time of return on the real time axis.}
\label{fig_SPcontrol}
\end{figure}

In this section we discuss how modifying the parameters of the driving fields can affect the degree of spin polarization of the recolliding electrons.
In particular, we analyze the effect of varying the relative intensities of the two counter-rotating fields.
Fig. \ref{fig_SPcontrol} contains a representation of the electric fields resulting from making the intensity of the second harmonic half and twice the intensity of the fundamental field (see parameters of the fields in fig. \ref{fig_SPcontrol} and in its footnote).
Increasing the relative intensity of the fundamental field shrinks the width of the field lobes.
Enhancing the relative intensity of the second harmonic has the opposite effect.
The corresponding recollision energy and spin polarization, obtained with these fields, are shown in \ref{fig_SPcontrol}, as a function of the recollision time, for one optical cycle of the fundamental field.
For comparison purposes, the results obtained for equal intensities of the counter-rotating fields (already discussed in the previous section), are included in fig. \ref{fig_SPcontrol}.

Spin polarization is presented in fig. \ref{fig_SPcontrol} (lower panels), also as a function of the recollision time.
We can see that relatively small modifications of the fields intensities lead to dramatic changes in the degree of polarization, allowing to achieve a high degree of control.
In particular, by tuning the relative intensities of the fields, it is possible to select the instant at which spin polarization changes it sign: increasing the intensity of the fundamental field shifts the change of sign towards earlier times, whereas increasing the intensity of its second harmonic has the opposite effect.

\section{Conclusions} \label{section_conclusions}

The possibility of inducing recollision with spin-polarized electrons can open new directions in attosecond spectroscopy.
Electron spin and orbital angular momentum can play an important role in well-established recollision-driven techniques such as photoelectron diffraction and holography \cite{SpannerJPB2004,MeckelScience2008,HuismansScience2011,BlagaNat2012,MeckelNatPhys2014,PullenNatComm2015} or high harmonic generation \cite{MilosevicPRA2000,MilosevicPRA2000_2,WornerScience2011,FleischeNatPhot2014,PisantyPRA2014,MedisauskasPRL2015,KfirNatPhot2015,CireasaNatPhys2015,SmirnovaJPB2015}.
We have shown that the use of intense two-color counter-rotating bi-circular fields can drive electron-core recollision with a degree of spin polarization that depends on the recollision time and therefore on the recollision energy.
Electron spin polarization upon tunnel ionization is intrinsically related to the generation of spin-polarized currents in the ionic core \cite{BarthJPB2014}.
In this context, the potential of inducing recollision within one optical cycle of the driving field can allow for probing spin-polarized currents in atoms and molecules with sub-femtosecond and sub-Angstrom resolution.
The time-dependence of spin polarization could be exploited to reconstruct information of the recollision process itself from spin-resolved measurements of diffracted electrons.
Furthermore, our work shows that the degree of spin-polarization can be modified as desired by tailoring the driving fields.
In particular, we have found that small variations in the relative intensities of the counter-rotating fields can change dramatically the level of polarization of the recolliding currents, opening the way for attosecond control of spin-resolved dynamics in atoms and molecules.

\section{Acknowledgements}

The authors acknowledge support from the DFG grant SM 292/2-3 and from the DFG SPP 1840 ``Quantum Dynamics in Tailored Intense Fields''.

\bibliography{Bibliography}

\end{document}